# Magnetic, electrical resistivity, heat capacity and thermopower anomalies in CeCuAs$_2$

Kausik Sengupta, E.V. Sampathkumaran[*] and S. Rayaprol
*Tata Institute of Fundamental Research, Homi Bhabha Road, Mumbai – 400005, India*

T. Nakano, M. Hedo, M. Abliz, N. Fujiwara and Y. Uwatoko
*The Institute for Solid State Physics, The University of Tokyo, 5-1-5 Kashiwanoha, Kashiwa, Chiba 277 8581, Japan*

K. Shigetoh and T. Takabatake
*Department of Quantum Matter, ADSM, Hiroshima University,
Higashi-Hiroshima 739-8530, Japan*

Th. Doert and J.PF. Jemetio
*Technische Universität Dresden, Institut für Anorganische Chemie, Mommsenstrasse 13, D-01062 Dresden, Germany.*

The results of magnetic susceptibility ($\chi$), electrical resistivity ($\rho$), heat-capacity (C) and thermopower (S) measurements on CeCuAs$_2$, forming in ZrCuSi$_2$-type tetragonal structure, are reported. Our investigations reveal that Ce is trivalent and there is no clear evidence for long range magnetic ordering down to 45 mK. The $\rho$ behavior is notable in the sense that (i) the temperature (T)-coefficient of $\rho$ is negative in the entire range of measurement (45 mK to 300 K) with large values of $\rho$, while S behavior is typical of metallic Kondo lattices, and (ii) $\rho$ is proportional to $T^{-0.6}$ at low temperatures, without any influence on the exponent by the application of a magnetic field, which does not seem to classify this compound into hither-to-known non-Fermi liquid (NFL) systems. In contrast to the logarithmic increase known for NFL systems, C/T measured down to 0.5 K exhibits a fall below 2 K. The observed properties of this compound are unusual among Ce systems.

PACS numbers: 71.27.+a, 72.15.Qm, 75.50.y, 72.15.-v

The search for novel Ce compounds remains unabated, and this direction of research got a boost after the discovery of non-Fermi-liquid (NFL) behavior [1,2] in the physical properties of Ce systems. In particular, there is a considerable interest in identifying NFL behavior in stoichiometric compounds. Recently, we have reported [3,4] unusual transport anomalies for 'normal' rare-earths (that is, other than Ce and Yb) for the series of compounds, RCuAs$_2$, crystallizing in ZrCuSi$_2$-type layered tetragonal structure [5-9]. Therefore, we considered it worthwhile to investigate the compound, CeCuAs$_2$, also taking note of the fact that the isostructural CeAgSb$_2$ and U compounds have been found to exhibit interesting properties [10-15]. Here, we report the results of magnetic susceptibility, $\chi$ (T= 1.8 – 300 K), electrical resistivity, $\rho$ (45mK – 300 K), thermopower, S (4.2 - 300 K), and heat capacity, C (0.5-30 K) measurements on CeCuAs$_2$. While we can not detect long range magnetic order in this compound, the observed characteristics are unusual, in particular, negative temperature coefficient of $\rho$ in the entire T-range of investigation (with large values of $\rho$ in the samples employed), with NFL-like characteristics at low temperatures without any corresponding anomaly in C, but showing a peak in C/T around 2K.

The sample CeCuAs$_2$ in the polycrystalline form was prepared as described in Ref. 4. Stoichiometric amounts of constituent elements (>99.9%) were sealed in an evacuated quartz tube and heated at 770 K for two days and 1170 K for 10 days. The specimen was then characterized by x-ray diffraction and there was no evidence for a secondary phase within the detection limit (2%). The lattice constants [$a$ = 4.045(1) and $c$= 10.111(2) Å] were found to be in good agreement with those reported in the literature [5]. For $\rho$ measurements, the sample was pressed into pellets and sintered again at 1170 K. The $\chi$ measurements were performed by a commercial magnetometer employing a magnetic field (H) of 5 kOe. The $\rho$ data were obtained by a conventional four-probe method in zero fields down to 1.4 K; the measurements were extended to 45 mK in a dilution refrigerator (Oxford Instruments) in the presence of 0, 10 and 40 kOe (below 7 K). The C data were collected in the range 0.5 - 10 K employing a commercial instrument (Quantum Design) and at higher temperatures by a homemade calorimeter. The S measurements (4.2 – 300 K) were performed by a differential method and the sample was suspended between two electrically isolated copper blocks across which a temperature gradient of 0.05-0.3 K was maintained.

The results of $\chi$ measurements are shown in Fig. 1. It is obvious from this figure that the plot of inverse $\chi$(T) is nearly linear above 150 K and the value of the effective moment ($\mu_{eff}$= 2.68$\mu_B$) obtained from the slope is very close to that expected for trivalent Ce ions. Consistent with this, the lattice constants follow lanthanide contraction,

---

[*] Corresponding author e-mail: sampath@tifr.res.in

smoothly interpolating between the values of La and other heavier rare earths in this series [5].

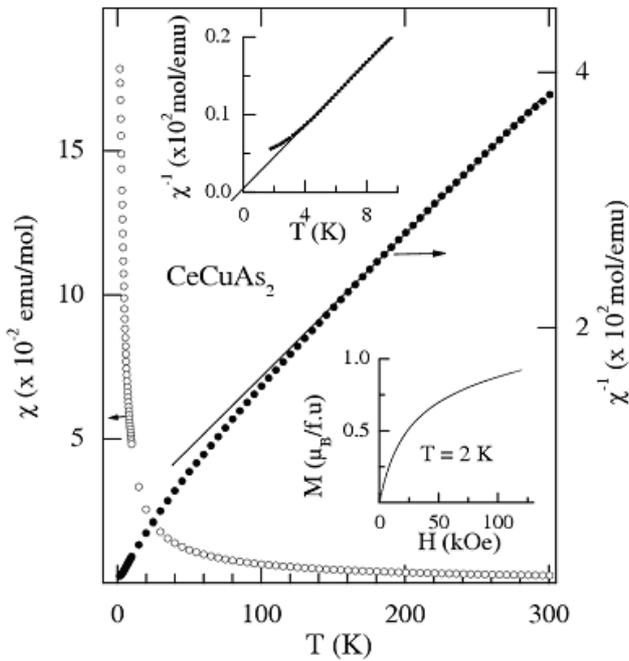

**FIG. 1.** Magnetic susceptibility ($\chi$) and inverse $\chi$ as a function of temperature for CeCuAs$_2$ and a line is drawn through the data points in the high temperature Curie-Weiss region. The $\chi^{-1}(T)$ at low temperatures is shown in an expanded form in an inset and a continuous line is drawn through the linear region. Another inset shows isothermal magnetization data at 2K

The value of the paramagnetic Curie temperature ($\theta_p$) in this temperature range is -47 K. As T is lowered, there is a gradual deviation of this plot from linearity, which is attributed to crystal-field effects. Since this compound forms in a tetragonal structure, it is expected that the crystal-field-split ground state is a doublet. It should be noted that there is no clear evidence for magnetic ordering down to 2 K (*vide infra*) and therefore the large value of $\theta_p$ with a negative sign mentioned above presumably arises from the Kondo interaction. The value of $\theta_p$ obtained from the low temperature data (say, below 10 K; see Fig. 1, top inset) however is close to zero and therefore the exchange interaction corresponding to the doublet ground state is very weak. We have also measured isothermal magnetization (M) at 2 K and we find (see Fig. 1, bottom inset) that M is non-hysteretic without any pronounced saturation even at 120 kOe, which rules out ferromagnetic ordering. Absence of hysteresis behavior rules out spin-glass freezing as well down to 2 K. The overall shape of M(H) curve is consistent with what one expects for the influence of H on the paramagnetic Ce ions at 2 K. The fact that the value of the magnetic moment is less than 1 $\mu_B$ even at very high fields is consistent with the proposal that the ground state is a crystal-field-split doublet.

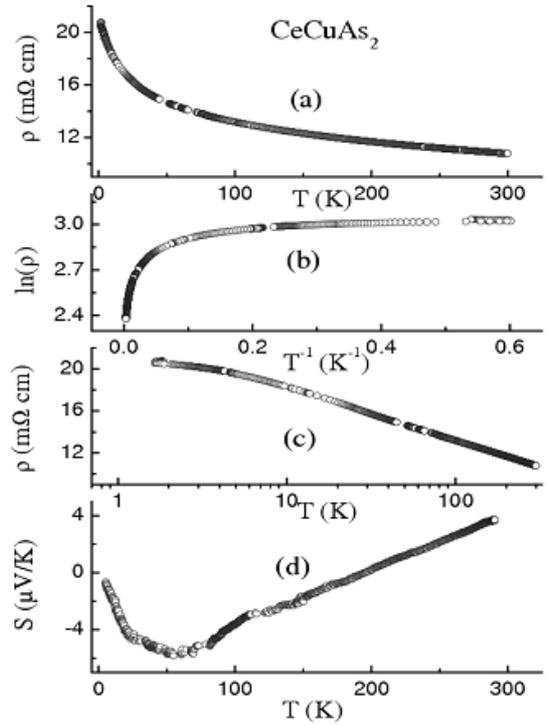

**FIG. 2.** Electrical resistivity ($\rho$) as a function of Temperature (1.5-300 K) for CeCuAs$_2$ shown in various ways: (a) Linear scale; (b) $ln\rho$ versus 1/T; (c) $\rho$ versus $log$T. The thermopower (S) data is plotted in (d).

We now discuss the $\rho$ data shown in Fig. 2a. It is to be noted that the value of $\rho$ is quite large at 300 K (of the order of 11 m$\Omega$ cm), which implies that this compound is strictly not a good metal. In addition, $\rho$ keeps increasing with decreasing T down to the lowest temperature measured (45mK). It is interesting that *this situation is completely different from that observed for all the other members of this rare-earth series [4]*; for instance, the non-magnetic analogues have been found to exhibit metallic behavior. Therefore, this behavior is intrinsic to Ce-4f electrons in this compound. Thus, the double-peaked structure due to interplay between crystal-field and the Kondo effects expected for trivalent Ce alloys [16] is absent for this compound. This prompts us to raise the question whether this compound could be classified as a Kondo semi-metal - a class of compounds attracting a lot of attention in the literature [see, for instance, Refs. 17-19]. In order to look for activated behavior over a wide T range, we have plotted $ln\rho$ as a function 1/T in Fig. 2b and it is apparent that the plot is not linear. Similar behavior has actually been reported for few other Ce-based Kondo semiconductors [18,19], e.g., Ce$_3$Bi$_4$Pt$_3$ [18], due to possible T-dependent gap effects. However, we find that $\rho$ varies essentially logarithmically (Fig. 2c) above 5 K, as though negative d$\rho$/dT arises from the single-ion Kondo effect. In order to clarify this issue further, we have measured S as a function of T, as the Kondo semi-conducting behavior should result in large values of S [20]. It is obvious from the figure 2d that the magnitude of S is considerably small (<4$\mu$V/K) compared to the values known [21] for Ce$_3$Bi$_4$Pt$_3$ in the T-range of investigation. The observed sign crossover in S(T) with a negative minimum is also characteristic of many of the trivalent-Ce Kondo lattices without gap effects [20]. Therefore, we tend to believe that, in $CeCuAs_2$, there is no energy-gap, at least above 4.2 K.

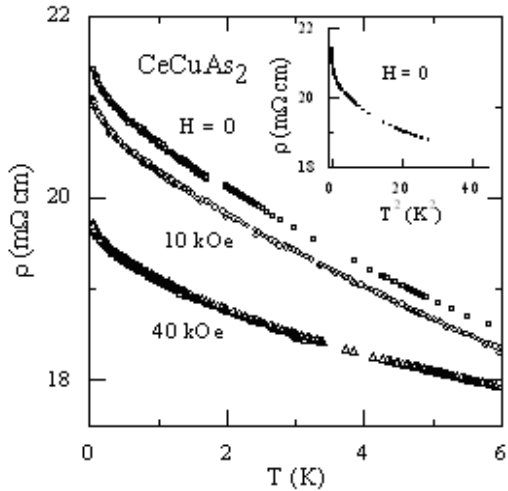

**FIG. 3.** Electrical resistivity ($\rho$) as a function of temperature in the range 45 mK to 6 K for $CeCuAs_2$ in the absence and in the presence of magnetic fields. In the inset, the zero-field data is plotted as a function of $T^2$.

We show the $\rho$ data from 45mK to 6 K in Fig. 3 at various fields. It is apparent that $\rho$ continues to increase monotonically with decreasing T, without any fall expected for coherent Kondo-lattices or magnetically ordering systems. It is to be noted that the $T^2$ dependence expected for the Kondo-related Fermi-liquid state is also absent (see Fig. 3, inset), thus showing NFL-like behavior. In this sense, this compound appears to belong to a small class stoichiometric compounds exhibiting NFL behavior (see, for instance, Refs. 22-24]. However, the magnitude (-0.6) of the exponent is much smaller than that what is usually seen (greater than 1) in other NFL systems. Another important finding is that an application of external field of 0 or 40 kOe (see Fig. 3) does not change the value of this exponent noticeably, though there is a reduction of the $\rho$ values (say, by about 8% for H= 40 kOe at 0.1 K). Thus, the observed NFL-like behavior is robust, in sharp contrast to the general expectation [25] that H should restore Fermi-liquid behavior (that is, exponent moving towards 2 with increasing H). It is difficult to explain the apparent NFL behavior in terms of Kondo-disorder models [26], since it is expected that the application of H should change the distribution of $T_K$ resulting in a variation of the exponent. Thus anomalously low value of the exponent, which is insensitive to H, coupled with large values of $\rho$, raise a question [27] whether effects other than what are usually described for previously known NFL systems have to be explored.

Above interpretation of the $\rho$ data in terms of NFL-like behavior implies that there is no magnetic ordering below 3 K in this material. Though we have sufficient evidence for this from isothermal M and $\chi$ data as well as from the (almost zero) low temperature $\theta_p$ value as described above, we have addressed this question microscopically by $^{63}Cu$ NMR measurements [28] in the range 0.7 to 300 K employing a pulse spectrometer by varying H at a fixed frequency of 33 MHz. Preliminary measurements clearly reveal the existence of Cu NMR signal even at 0.7 K, which would not be the case, if the compound is magnetically long-range-ordered (that is, magnetic periodicity). The magnitude of the Knight shift keeps increasing with decreasing T, attaining a value of -3.2% around 2 K. We have also attempted to measure spin-lattice relaxation time, but there are difficulties in obtaining absolute values due to complex spin-echo signal at low temperatures due to increasing broadening of the signal with decreasing T. Hence, at present, it is difficult ascertain from these relaxation rate studies whether there are energy-gaps developing below 3 K as a possible 'other effect' mentioned in the previous paragraph.

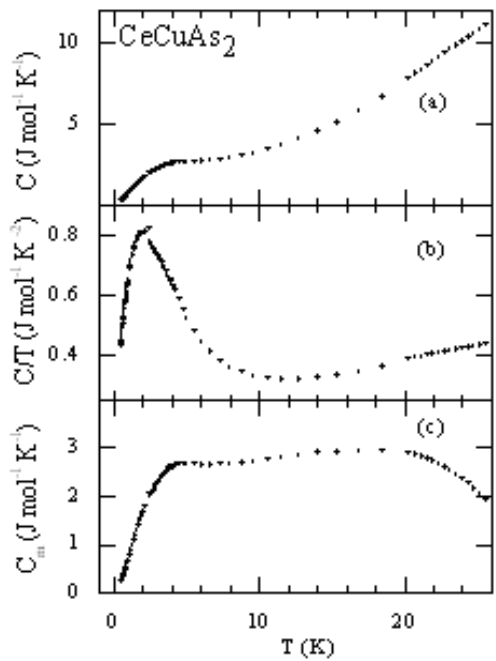

**FIG. 4.** The plots of (a) Heat-capacity (C) as a function of temperature (0.5 to 25 K), (b) C/T versus T, and (c) the 4f contribution ($C_m$) to C, for $CeCuAs_2$.

In order to look for other anomalies known for NFL systems, we have performed C measurements. (The $\chi$ could be measured down to 2 K only and hence we would not like to discuss functional dependence for this property). There is no distinct $\lambda$-anomaly or a peak in the plot of in C(T) (Fig. 4a), characteristic of long-range magnetic order of a ferromagnetic-type or an anti-ferromagnetic-type. However, there is a gradual fall below 4.5 K. This feature is more clearly visible (Fig. 4b) in the plot of C/T, in which a peak is seen at about 2 K. It is difficult to attribute this to long range ordering, as the fall above the peak is rather sluggish; that is, it persists till about 8 K. For a comparison, the reader may see the plots of C/T versus T for other isostructural Ce compounds undergoing long-range magnetic order (Ref. 12). Such a broad feature may not be attributed to spin-glass freezing, considering that (i) there is no hysteretic behavior in isothermal M at 2 K (as mentioned earlier), (ii) we do not find any peak even in the low field (100 Oe) $\chi$ data (not shown here) with a bifurcation of zero-field-cooled and field-cooled χ(T) curves around 4 K. Therefore, the drop in C below 4 K may have a different origin. If this compound is a non-magnetic heavy-fermion, from the χ value at 2 K (assuming Wilson ratio of 1), one can expect that the value of C/T should be much larger than 2J/molK$^2$ at the lowest temperature. The gradual rise of C/T below 8 K may imply a tendency towards such a large value at much lower temperatures (with already a large value of about 800mJ/molK$^2$ at the peak), but it is cut off by the drop resulting in a value of about 425 mJ/molK$^2$ at 0.5 K. We have also obtained (Fig. 4c) the 4f-contribution (C$_m$) to C from the knowledge [29] of the C values of YCuAs$_2$ (Ref. 4) and the 4.5K-peak can not be fitted to a Schottky anomaly arising from crystal-field splitting. In any case, the crystal-field splitting between the ground state and the first excited state should be of a much larger magnitude, compared to the peak temperature of 4.5 K, judged by the corresponding value in the isostructural compound, CeAgSb$_2$ [Ref. 11] (We believe that the broad peak in C$_m$(T) around 20 K may arise from the crystal-field effects; but we would like to view this feature with some caution, as one is not sure of the subtraction procedure for phonon contribution at such high temperatures). Therefore, the low temperature drop in C/T is puzzling, as one would have expected a rise (generally logarithmically) with decreasing T if this system has to be classified with other known NFL systems. One possible way to explain this anomaly is to propose that there is a disorder-driven continuous metal-insulator transition in some clusters of the specimen, resulting in a gradual loss of density of states with decreasing T. Such a concept has actually been invoked as a mechanism of NFL behavior [30], but the low-temperature downturn in C/T is not predicted by the theory of Ref. 30; it is of interest to explore a modification of this theory to find conditions under which a drop in C/T is expected. In short, there is no straightforward explanation for the C anomaly at the moment [31].

To conclude, the stoichiometric trivalent-Ce compound, CeCuAs$_2$, exhibits negative dρ/dT down to 45 mK, with a NFL-like behavior at low temperatures without showing a corresponding anomaly in C. On the one hand, S(T) behavior, ruling out possible *semi-conducting nature* as a cause of negative dρ/dT above 5 K, is consistent with the trends among metallic *Kondo lattices* of Ce; the observed logarithmic T-dependence of ρ above 5 K seems to support an interpretation in terms of the Kondo effect for the negative dρ/dT. On the other hand, it is interesting that double-peaked structure known for the trivalent metallic Ce-based Kondo lattices due to an interplay with the crystal-field effects is absent in the ρ(T) data, despite indications for the crystal-field effects, say, in the magnetization data. There is an urgent need therefore to perform inelastic neutron scattering studies to understand crystal-field scheme. The absolute values of ρ are large and, if one can rule out possible role of grain boundary effects on this large magnitude by studies on single crystals, this compound could serve as a model system for theoretical investigations for the Kondo behavior in the large limit of ρ. Given that the compound does not undergo long-range magnetic ordering down to 45 mK, the small value of the exponent of the T-dependence of ρ for T<5 K and its insensitivity to applied fields and the absence of a corresponding rise in C/T (but showing a decrease below 2 K) with decreasing T make this compound different from other Ce compounds. It will be of interest to focus future studies to explore the existence of novel gaps [30] or even localization [32] effects due to weak-disorder as well as other novel magnetic effects to understand the low temperature anomalies. Finally, we would like to mention that the properties of C and ρ (measured down to 0.5 K) as well as of χ and M, of the composition, CeCu$_{1.1}$As$_2$, the 'stuffed variant' of ZrCuSi$_2$ structure [8], are similar to the stoichiometric compound, offering overall confidence to the findings reported in this article; however, ρ varies linearly below 5 K indicating possible small variations of the exponent value depending on the stoichiometry.

[1]C.L. Seaman et al., Phys. Rev. Lett. **67,** 2882 (1991).
[2]See, for a review, H.v. Löhneyson, J. Magn. Magn. Mater. 200, 532 (1999).
[3]E.V. Sampathkumaran, K. Sengupta, S. Rayaprol, K.K. Iyer, Th. Doert and J.P.F. Jemetio, Phys. Rev. Lett. **91,** 036603 (2003).
[4]Kausik Sengupta, S. Rayaprol, E.V. Sampathkumaran, Th. Doert and J.P.F. Jemetio, Physica B, in press (see arXiv cond-mat: 0401599).
[5]M. Brylak, M.H. M\"oller, and W. Jeitschko, J. Solid State Chem. **115**, 305 (1995).
[6]M. Wang, R. McDonald, and A. Mar, J. Solid State Chem. **147**, 140 (1999).
[7]Y. Mozharivskyj, D. Kaczorowski, and H.F. Franzen, J. Solid State Chem. **155**, 259 (2000).
[8]J.-P. Jemetio, Th. Doert, O. Rademacher and P. Böttcher, J. Alloys and Comp. **338,** 93 (2002).
[9]J.-P. Jemetio et al, Z. Kristallogr. NCS **217**, 445 (2002).
[10]K.D. Myers, S.L. Bud'ko, I.R. Fisher, Z. Islam, H. Kleinke, A.H. Lacerda, and P.C. Canfield, J. Magn. Magn. Mater. **205,** 27 (1999)
[11]S. Araki, N. Metoki, A. Galatanu, E. Yamamoto, A. Thamizhavel, and Y. Onuki, Phys. Rev. B **68,** 024408 (2003).
[12]A. Thamizhavel, T. Takeuchi, T. Okubo, M. Yamada, R. Asai, S. Kirita, A.Galatanu, E. Yamamoto, T. Ebihara, Y. Inada, R. Settai, and Y. Onuki, Phys. Rev. B **68,** 054427 (2003).
[13]T. Takeuchi, A. Thamizhavel, T. Okubo, M. Yamada, N. Nakamura, T. Yamamoto, Y. Inada, K. Sugiyama, A. Galatanu, E. Yamamoto, K. Kindo, T. Ebihara, and Y. Onuki, Phys. Rev. B **67,** 064403 (2003).
[14]V.A. Sidorov, E.D. Bauer, N.A. Frederick, J.R. Jeffries, S. Nakatsuji, N.O. Moreno, J.D. Thompson, M.B. Maple, and Z. Fisk, Phys. Rev. B **67,** 224419 (2003).
[15]D. Kaczorowski, R. Kruk, J.P. Sanchez, B. Malaman, and F. Wastin, Phys. Rev. B **58**, 9227 (1998).
[16]B. Cornut and B. Coqblin, Phys. Rev. B **5**, 4541 (1972).
[17]S. Nishigori, H. Goshima, T. Suzuki, T. Fujita, G. Nakamoto, H. Tanaka, T. Takabatake, and H. Fujii, J. Phys. Soc. Japan **65,** 2614 (1996).
[18]M.F. Hundley, P.C. Canfield, J.D. Thompson, Z. Fisk

[18] and J.M. Lawrence, Phys. Rev. B **42**, 6842 (1990).

[19] E.D. Bauer, A. Slebarski, E.J. Freeman, C. Sirvent and M.B. Maple, J. Phys.: Condens. Matter **13**, 4495 (2001).

[20] T. Takabatake, T. Sasakawa, J. Kitagawa, T. Suemitsu, Y. Echizen, K. Umeo, M. Sera and Y. Bando, Physica B **328,** 53 (2003) and references therein**.**

[21] K. Katoh and T. Takabatake, J. Alloys and Compounds **268,** 22 (1998).

[22] P. Gegenwart, F. Kromer, M. Lang, G. Sparn, C. Geibel and F. Steglich, Phys. Rev. Lett. **82**, 1293 (1999).

[23] F.M. Grosche, S.R. Julian, N.D. Mathur and G.G. Lonzarich, Physica B 223-224, 50 (1996).

[24] D. Kaczorowski, O. Tougait, A. Czopnik, Cz. Marucha and H. Noel, J. Magn. Magn. Mater. (2003), in press.

[25] See, for instance, J. Paglione, M.A. Tanatar, D.G. Hawthorn, E. Boaknin, R.W. Hill, F. Ronning, M. Sutherland, L. Taillefer, C. Petrovic and P.C. Canfield, Phy. Rev. Lett. **91**, 246405 (2003).

[26] E. Miranda et al, Phys. Rev. Lett. **78,** 290 (1997).

[27] J.S. Kim, E.-W. Scheidt, D. Mixson, B. Andraka and G.R. Stewart, Phys. Rev. B **67**, 184401 (2003).

[28] N. Fujiwara et al (to be published)

[29] To derive $C_m$, we have followed the procedure, prescribed by J.A. Blanco et al Phys. Rev. B 43, 13145 (1991). We further derived the magnetic entropy ($S_m$) as a function of temperature. The value of $S_m$ is about 60% of R$ln$2 at 4 K.

[30] E. Miranda and V. Dobrosavljevic, Phys. Rev. Lett. **86,** 264 (2001).

[31] M. Ishikawa, N. Takeda, M. Koeda, M. Hedo and Y. Uwatoko, Phys. Rev. B **68**, 024522 (2003). This article has reported a similar C/T peak for some compositions in the Ce-Cu-Si system in the presence of magnetic fields (after superconductivity is destroyed) and the exact origin is still unclear.

[32] Since the exponent value below 5 K is closer to that expected (0.5) for weak localization effects [See, P.A. Lee and T.V. Ramakrishnan, Rev. Mod. Phys. **57**, 287 (1985)], we are tempted to think along these lines.